\documentclass[showpacs,preprintnumbers,amsmath,amssymb,aps]{revtex4}
\usepackage{graphics}

\begin{document}

\title{Parametrization of the equation of state and the expanding universe}
\author{
W. F. Kao\thanks{%
wfgore@cc.nctu.edu.tw} \\
Institute of Physics, Chiao Tung University, Hsinchu, Taiwan}

\begin{abstract}
The structure of the equation of state $\omega$ could be very
complicate in nature while a few linear models have been
successful in cosmological predictions. Linear models are treated
as leading approximation of a complete Taylor series in this
paper. If the power series converges quickly, one can freely
truncate the series order by order. Detailed convergent analysis
on the choices of the expansion parameters is presented in this
paper. The related power series for the energy density function,
the Hubble parameter and related physical quantities of interest
are also computed in this paper.
\end{abstract}

\maketitle

\section{INTRODUCTION}
Recently there have been advances in our abilities in cosmological
observations for the quest of exploring the expansion history of
the universe. It carries cosmology well beyond determining the
present dimensionless density of matter $\Omega_m$ and
deceleration parameter $q_ 0$ \cite{[1]}. It seems possible to
reconstruct the entire function $a(t)$ representing the expansion
history of the entire universe. Earlier on, cosmologists sought
only a local measurement of the first two derivatives of the scale
factor a, evaluated at a single time $t_ 0$. Attempt will be made
trying instead to map out the function determining the global
dynamics of the universe in the near future. One notes that many
qualitative elements of cosmology follow directly from the
structure of the metric\cite{[2]}. Therefore, deeper understanding
of our universe requires knowledge of the dynamics of the scale
factor $a(t)$ echoing the transition of energy between components
from the epoch of radiation domination to that of matter
domination. This is also known to be a key element in the growth
of density perturbations into large structure. Indeed, many
proposed observation will be able to probe the function a(t) more
completely throughout all ages of the universe
\cite{[3],[4],[5],[6]}.

A number of promising methods are being developed including the
magnitude-redshift relation of Type Ia supernovae. The goal of
mapping out the recent expansion history of the universe is well
motivated. The thermal history of the universe, extending back
through structure formation, matter-radiation decoupling,
radiation thermalization, primordial nucleosynthesis, etc. is very
important in the study of cosmology and particle physics, high
energy physics, neutrino physics, gravitational physics, nuclear
physics, and so on \cite{[7]}.

Expansion history of the universe is similarly a very promising
research focus with the discovery of the current acceleration of
the expansion of the universe. This includes the study of the role
of high energy field theories in the form of possible
quintessence, scalar-tensor gravitation, higher dimension
theories, brane worlds, etc in the very recent universe. The
accelerated expansion is also important to the possible fate of
the universe \cite{[8],[9],[10]}. It is hence important to obtain
the magnitude-redshift law relating to the scale factor-time
behavior $a(t)$ from these supernova observations with the
proposed Supernova/Acceleration Probe mission \cite{[11]}.

Therefore, the study of modelling different equation of state
(EOS) derived from different theories plays an important role in
the study of the recent expansion history of our universe. The
study of dark matter and possible contribution from different
combinations of different theories including the SUGRA and string
theory will become important also in the near future once the LHC
experiment starts. Therefore all practical methods to study its
impact on the evolution history of our universe deserve full
attentions. In fact, there have been proposals to parameterize
different candidate models that are helpful in extracting
important physics with the help of several different linear
parameterizations of parameter $z/(1+z)$ or $z$ for the
corresponding equation of state function $\omega$
\cite{[12],[13],[14],[15],[17],[18],[19],[20],[21]}. These linear
models are shown to be very successful in recreating various
physical functions.

Linear models, e.g. $\omega=\omega_0+\omega_1 y$ with $y=z/(z+1)$,
should be thought of as leading approximation of a complete Taylor
series. If this is so, result derived from higher order in $y$,
e.g. $O(y^2)$ terms, should only provide marginal contribution and
hence can be ignored. If the higher order contribution is
appreciable, one should be very careful in dealing with the
truncated series. As a practical analysis, one should compute all
related physical functions also as a power series order by order
to obtain reliable observable according to the precision
requirement. Otherwise, unphysical contributions may build up and
leave the final result invalid. Fortunately, this complication may
be unnecessary if the higher order contributions are small as
compared to the leading order contribution.

Therefore, we try to extend the parametrization further to include
the effect of higher orders of the series expansion results.
Related physical quantities are treated carefully order by order
in order to extract more reliable information from these
expansions. One also tries to determine the expansion coefficients
from the fitting of the measured Hubble parameters or energy
density. As a result, one may reconstruct the series expansion of
the equation of state and probe the nature and origin of the
matter sources.

Finally, we will also present an error analysis to find the range
of convergence and possible error control for a meaningful
truncation. For example, the linear model given by
$\omega=-0.82+0.58 y$ fits the SUGRA result to a good precision at
large $z$ even the linear model is already off $27 \%$ when $z
\sim 1.7$\cite{[15]}. One can show that the next leading expansion
coefficient with $\omega_2y^2 <0.1y^2$ only provides at most $20
\%$ error even when the redshift is close to 1100. As a result,
one can show that a reliable truncation is possible when the
higher order expansion coefficient is small for all range
$y=z/(z+1)=0 \to 0.9991$ up to the last scattering surface when $z
\sim 1100$. Therefore, it does not matter much whether one sums up
all power of $y$ coupled to the leading coefficients $\omega_0$
and $\omega_1$. In addition, the linear model $\omega=-0.4+0.11 y$
approximates the inverse power law model very well at small
$y$\cite{[15]}. Similar analysis will also be shown in this paper
for comparison.

In addition, the evolution of the equation of state $\omega$ of
the physical universe could be very complicate. Indeed, the
equation of state can be described by $\omega=p/\rho=(\dot\phi^2/2
-V)/(\dot\phi^2/2 +V)$ for many different effective theory with
effective potential $V$ and scalar field $\phi$. For example,
$V_{INV} \sim \phi^{- \alpha}$ and $V_{SUGRA} \sim \phi^{- \alpha}
\exp [\phi^2]$ for inverse power law models and SUGRA models
respectively\cite{[15]}. The nature of our physical universe may
also consist of many other combinations of scalar fields
contributions. It is therefore important for us to figure out
which model plays the most important role in the expansion history
of our universe. The linear model or leading approximation appears
to be of help in determining the essential part of the nature of
the equation of state and its corresponding effective theory.
Therefore a more detailed analysis on the convergence properties
of various approximated models is very important as a research
topic.

 Since one is not sure about the nature of the function
$\omega$, the Taylor series expanding $\omega(y)=\sum_n
\omega_ny^n$ into series sum with expansion parameter $\omega_n$
may also be used to determine leading expansion parameter with the
help of future measurement. By fitting the Hubble parameter
measurements one could determine leading terms of the expansion
coefficients $\omega_n$. As a result, one would be able to
reconstruct the recent evolution of the function $\omega$ to the
better precision. We will present the complete analysis in this
paper. Since future observations will be able to provide tight
constraint for only a few parameters in the near future, the
material provides in this paper can be used to determine these
parameters in a more reliable details. Hopefully, future
development of observation tools may provide deeper insight to
determine more constraints in these theories.

We will present, in section II, a Taylor series expansion of some
physical models. One will explain the choices of the expansion
variables which come from the integration involving $dz/(1+z)$.
The convergence range of these expansion variables will also be
presented. In section III, we will also compute leading terms of
the energy density function $\rho$, the Hubble parameter $H$, and
the conformal time $\eta$ for these expansions in order to discuss
the difference between these approaches.(cf. \cite{[14]} and Fig.
1 of \cite{[15]}). Due to the fact that the parameter $z/(1+z)$
has a larger naive convergent range, this parameter could be more
useful in large $z$ expansion. We also compute and analyze the
convergence range of a linear model in section IV.

 \section{Taylor Series of the Equation of State}
The equation of state is defined by the relation $\omega =
p/\rho$. In addition, the field equations of the universe can be
shown to be:
\begin{eqnarray} \label{eos}
d \ln \rho (z)  &=& 3 (\omega +1) d \ln (1+z), \\
H^2 &=& {8 \pi G \over 3} \rho, \label{H}
\end{eqnarray}
for a flat FRW universe with $1+z$ the redshift defined via $a/a_0
= 1/(1+z)$. Various linear models \cite{[12]}, \cite{[13]} have
been suggested for the EOS in the literatures. We will try to
point out in this paper the relations between these linear models
and go further to obtain a more complete leading-order expansions.

Taylor expansion is known to be one of the best ways to extract
leading contributions from a generic theory with the help of a
power series expansion of some suitable field variables. The
Taylor series is normally convergent quickly depending on the
structure of the expansion coefficients. The power series will,
however, converge quickly if the range of variable is properly
chosen. Hence it is important to choose an appropriate expansion
variable for the purpose of our study.

For example, one may expand the EOS, assuming to be a smooth
function for all $z$, as a power series of the variable $z$ around
the point $z=0$. This will lead to the power series expansion:
\begin{equation}\label{z}
\omega(z)= \sum_{n =0}^{\infty} {\omega^{(n)}(z=0) \over n!} z^n .
\end{equation}
Here the summation of $n$ runs from $0$ to $\infty$ and
$f^{(n)}(z)={d^n f(z) / dz^n}$ for any function $f(z)$. From now
on, the range of summation will be omitted for convenience
throughout this paper unless it is different from the range from
$0$ to $\infty$.

One sees clearly that there are three useful and practical ways to
expand the equation of state: $z$, $1/(1+z)$, and $\ln (1+z)$,
from the structure of the conservation law given by Eq.
(\ref{eos}). It turns out that the expansion parameters will
become $z$, $-z/(1+z)$, and $\ln (1+z)$ respectively if one tends
to expand the function $\omega$ about the point $z=0$. In
practical, we will expand the second choice with the parameter
$z/(1+z)$ instead of $-z/(1+z)$ for convenience.

In this section, one will first expand the EOS as a power series
of the variable $w=1/(1+z)=a/a_0$ around the point $w(z=0)=1$. The
result is
\begin{equation} \label{wz}
\omega(w(z))= \sum_n {\omega^{(n)}(w=1) \over n!} (w-1)^n = \sum_n
{\omega^{(n)}(w=1) \over n!} (-)^n ({z \over 1+z})^n .
\end{equation}

Note that if we expand the EOS as a power series of the variable
$y=z/(1+z)$ around $y(z=0)=0$, we will end up with a similar power
series:
\begin{equation} \label{yz}
\omega(y(z))= \sum_n {\omega^{(n)}(y=0) \over n!} y^n .
\end{equation}

The Taylor series are normally convergent quickly depending on the
structure of the expansion coefficients. Nonetheless, one would
prefer to choose a more appropriate expansion parameter in order
to make the series converge more rapidly. As a result, leading
order terms will be enough to extract the most important physics
from the underlying theory. Therefore the advantage of the $y$
expansion is that the power series converges rapidly for all range
of the parameter $-1< y=z/(1+z) <1$, or equivalently, $-1/2<z<
\infty$ as compared to the range $|z|<1$ for the power series
expansion of $z$ shown in Eq. (\ref{z}).

One notes, however, that one should also expand all physical
quantities and the field equations to the same order of precision
we adopted for the EOS expansion. Higher order contributions will
not be reliable unless one can show that the higher order terms
does not affect the physics very much. For example, the liner
order is good enough for the expansion of the EOS modelling SUGRA
model \cite{[13]}. This is because that the linear term fits the
predicted EOS for SUGRA model to a very high precision. One
readily realizes, however, that it is not easy to track the series
expansion order by order due to the form of the Eq. (\ref{eos})
for the EOS for the $y$ expansion. This is because that after
performing the integration, one needs to pay attention to the
distorted integration result. Indeed, one needs to write $y=1-w$
in order to perform the integration involving $d \ln w$. Indeed,
one will need to recombine the result back to a power series of
$y$. The trouble is that the lower order terms could sometimes
hide in the higher order terms in $w$. It may not be easy to track
clearly the lower order $y$ expansion when we perform the
expansion, for example, due to the exponential factor in the Eq.
(\ref{eos}). Therefore, one finds that the most natural way to
expand the EOS is to expand it in terms of the variable $x= \ln
(1+z)$ around the point $x(z=0)=0$. Indeed, this power series can
be shown to be:
\begin{equation} \label{x}
\omega(x(z))= \sum_n {\omega^{(n)}(0) \over n!} x^n.
\end{equation}
And this power series converge very rapidly in the range $-1< \ln
(1+z) <1$, or equivalently, $-0.63 \sim 1/e - 1 < z <e -1 \sim
1.72$. Here $e \sim 2.72 $ is the natural factor. Note that this
limit happens to agree with the proposed scope of the SNAP
mission. We will study these different power series expansion for
the EOS and its applications in the following sections.

\section{Some practical expansions}

Due to the structure of the differential equations (\ref{eos}),
one finds that it will be easy for us to compute related Taylor
series by expanding $\omega$ either as functions of $z$, or $y
\equiv z/(1+z)$, or $x \equiv \ln z$. Therefore, we will present
details of these expansion series in this section. For
convenience, we will use repeated notation for the expansion
series in this section for convenience and economics of notations.
One should bear in mind that these coefficients are defined
differently associated with different arguments $y$, $x$ and $z$
defined in each subsection.
\subsection{Power series of $z/(1+z)$}

Note that the expansion parameter $y \equiv z/(1+z)= 1 -a/a_0$.
Therefore, the Taylor series expansion around small $y$ is
equivalent to a series expansion around $a=a_0$. This is again a
series expansion around the very recent universe near $a=a_0$. The
expansion series in $y$ has a naive convergence range for all $y
<1$ which is defined to be the early universe with $a < a_0$. We
will show in details how to extract the leading terms in the
$y$-expansion with $y={z / (1+z})$ around the point $y(z=0)=0$.
One can write the expansion coefficient as $\omega_n =
{(1+\omega)^{(n)}(y=0) / n!}$ such that the power series for the
expansion of the EOS becomes
\begin{equation} \label{y}
1+\omega(y(z))= \sum_n \omega_n y^n .
\end{equation}
Hence the Eq. (\ref{eos}) can be shown to be
\begin{eqnarray} \label{eos1}
d \ln \rho (y)  = 3 (1+\omega)  { d y \over 1-y}= d\,\, \left[ \sum_n \sum_k 3
\omega_n {y^{n+k+1} \over n+k+1} \right ] .
\end{eqnarray}
Note that we are now expanding with respect to the smooth function
$(1+\omega)$, instead of $\omega$, as a Taylor series for
convenience. Therefore, one can integrate above equation to obtain
\begin{equation} \label{eos2}
\rho (y) = \rho_0 \exp [\,\, 3 \sum_n \sum_k {\omega_n y^{n+k+1}
\over n+k+1}\,\,] \equiv \rho_0 X(y)= \rho_0 \sum_n X_n y^n.
\end{equation}
Note that one needs to expand the function $\rho$ as a power
series of $y$ too in order to extract the approximated solution
with appropriate order. The expansion coefficient $X_n$ is defined
as $X_n = X^{(n)}(y=0)/n!$. Here, the superscript in $X'$ denotes
differentiation with respect to the argument $y$ of the function
$X(y)$. One can show that $X'=XY$ with $Y=3 \sum_n \sum_k \omega_n
y^{n+k}$. In addition, one can show that
\begin{equation}\label{Yl0}
Y^{(l)}(y)=3 \,\, \sum_{n} \sum_{k}{ (n+k)! \over (n+k-l)!}
\omega_n y^{n+k-l}.
\end{equation}
Hence one has
\begin{equation}\label{Yl}
Y^{(l)}(y=0)=3(l!) \,\, \sum_{n=0}^{l} \omega_n .
\end{equation}

One can also show, for example, that
\begin{eqnarray}
X''&=& X(Y^2+Y') \\
X'''&=& X(Y^3+3YY'+Y'') \\
X^{(4)} &=& X(Y^4+6Y^2Y'+3(Y')^2+4 YY''+Y''' ).
\end{eqnarray}
This series does not appear to have a more compact close form for
the multiple differentiation with respect to $y$. One can,
however, put the equations as a more compact format:
\begin{equation}
X^{(l+1)}=X [Y+{d \over dy}]^l Y.
\end{equation}
It appears, however, that one needs to do it manually even it is
straightforward. We will only list the leading terms as this is
already suitable expansion for our purpose at this moment.

Hence one has
\begin{eqnarray}
X_0&=&1, \\
X_1&=& 3 \omega_0, \\
X_2&=&{1 \over 2} [9 \omega_0^2+3\omega_0+3
\omega_1], \\
X_3&=& {1 \over 2} [9 \omega_0^3+9 \omega_0 (\omega_0+\omega_1)
+ 2 (\omega_0 + \omega_1 + \omega_2) ], \\
X_4 &=& {1 \over 8} [27 \omega_0^4+54 \omega_0^2
(\omega_0+\omega_1) + 9 (\omega_0 + \omega_1)^2 +
24\omega_0(\omega_0 + \omega_1 + \omega_2)+ 6(\omega_0 + \omega_1
+ \omega_2 +\omega_3) ].
\end{eqnarray}
Therefore, one can expand the final expression for the energy
density $\rho$ accordingly. Indeed, the result is
\begin{eqnarray}
\rho&=&\rho_0 \{  1 +  3 \omega_0 y + {1 \over 2} [9
\omega_0^2+3\omega_0+3 \omega_1] y^2 +  {1 \over 2} [9
\omega_0^3+9 \omega_0 (\omega_0+\omega_1)
+ 2 (\omega_0 + \omega_1 + \omega_2) ] y^3  \nonumber \\
&+& {1 \over 8} [27 \omega_0^4+54 \omega_0^2 (\omega_0+\omega_1) +
9 (\omega_0 + \omega_1)^2 + 24\omega_0(\omega_0 + \omega_1 +
\omega_2)+ 6(\omega_0 + \omega_1 + \omega_2 +\omega_3) ]y^4  \}
+O(y^5) \nonumber \\
&=& \rho_0 \{ \exp [3 \omega_0 y] + {3 \over 2} [\omega_0+
\omega_1] y^2 +  {1 \over 2} [9 \omega_0 (\omega_0+\omega_1)
+ 2 (\omega_0 + \omega_1 + \omega_2) ] y^3  \nonumber \\
&+& {1 \over 8} [54 \omega_0^2 (\omega_0+\omega_1) + 9 (\omega_0 +
\omega_1)^2 + 24\omega_0(\omega_0 + \omega_1 + \omega_2)+
6(\omega_0 + \omega_1 + \omega_2 +\omega_3) ]y^4  \} +O(y^5)
\end{eqnarray}
to the order of $y^4$. Note that one keep the order of precision
to $y^4$ in computing the energy density $\rho$ even we are
expanding the EOS only to the order of $y^3$. This is due to the
special structure in the energy momentum conservation law
(\ref{eos}). In addition, the largest power terms of $\omega_0$ in
the series come from the combination $\sum_n [Y^n(0)/n!]y^n=\exp[3
\omega_0 y]$ can be summed over directly. Moreover, one can also
show that the Hubble parameter $H=H_0X^{1/2}$ with $H_0=\sqrt{8\pi
G \rho_0 /3}$. And the expansion for $X^{1/2}$ can be obtained by
replacing all $\omega_n$ with $\omega_n/2$ in writing the
expansion for $X$. Therefore one has
\begin{eqnarray}
H&=&H_0  \{  \exp[{3 \over 2} \omega_0 y] + {3 \over 4} (\omega_0+
\omega_1) y^2 +  {1 \over 8} [9 \omega_0 (\omega_0+\omega_1)
+ 4 (\omega_0 + \omega_1 + \omega_2) ] y^3 + {1 \over 32}  [  \nonumber \\
&& 27 \omega_0^2 (\omega_0+\omega_1) +9 (\omega_0 + \omega_1)^2 +
24\omega_0(\omega_0 + \omega_1 + \omega_2)+ 12(\omega_0 + \omega_1
+ \omega_2 +\omega_3)  ] y^4  \} +O(y^5).
\end{eqnarray}

Note also that one can also compute the conformal time according
to the expression:
\begin{equation}
H_0 \eta = \int_0^z dz'X^{-{1 \over 2}}=\int_0^y dy' {X^{-{1 \over
2}} \over (1-y')^2 }
\end{equation}
which comes from the definition $d \eta =dt/a$. Knowing that
$1/(1-y)^2=\sum_n (n+1)y^n$, one can show that
\begin{equation}
H_0 \eta = \int_0^y dy' X^{-{1 \over 2}} \sum_n (n+1)y'^n .
\end{equation}
One can write $X^{-1/2}=\sum_k  x_k y^k= \sum_k X_k(\omega_n \to
-\omega_n/2) y^k$, and expand $H_0 \eta = \sum_k \eta_k y^k$ for
convenience. Therefore, one has
\begin{equation}
\eta_l= \sum_{n=0}^{l-1} {n+1 \over l} x_{l-n-1}
\end{equation}
for $l \ge 1$. Note that $\eta_0=0$.  As a result, one can easily
reconstruct the power series of $H_0 \eta$. Therefore, one has
\begin{eqnarray}
H_0 \eta &=&  y -[ {3 \over 4} \omega_0 -1]y^2 + [
 {3 \over 8} \omega_0^2 -{1 \over 4} \omega_1 -{5 \over 4} \omega_0 + 1]y^3  
 -[{9 \over 64} \omega_0^3- {27 \over 32} \omega_0^2  -{9 \over
32} \omega_0 \omega_1 +{13 \over 8} \omega_0 +{1 \over 2} \omega_1
+ {1 \over 8} \omega_2-1 ]y^4  +O(y^5) \nonumber \\
 &=& {y \over 1-y} - {3 \over 4} \omega_0 y^2 + [
 {3 \over 8} \omega_0^2 -{1 \over 4} \omega_1 -{5 \over 4} \omega_0 ]y^3
 -[{9 \over 64} \omega_0^3- {27 \over 32} \omega_0^2  -{9 \over
32} \omega_0 \omega_1 +{13 \over 8} \omega_0 +{1 \over 2} \omega_1
+ {1 \over 8} \omega_2 ]y^4 +O(y^5).
\end{eqnarray}
Note that terms independent of $\omega_0$, $y+y^2+ \cdots$, comes
from the combination $K_0 \equiv \sum_{l=1}^{\infty} \eta_l y^l =
\sum_{l=1}^{\infty} \sum_n^{l-1} (n+1)x_{l-n-1} y^l/l$. Taking the
terms with $n=l-1$ in $n$-summation, one can show that this
summation becomes $\sum_{l=1}^{\infty} x_0 y^l=
\sum_{l=1}^{\infty} y^l =y/(1-y)$.

\subsection{power series of $\ln (1+z) $}

Note that the expansion parameter $x \equiv \ln (1+z)= - \ln (
a/a_0)$. Therefore, the Taylor series expansion around small $|x|$
is equivalent to a series around $a=a_0$. This is again a series
expansion around the very recent universe near $a=a_0$. Indeed,
the expansion series in $x$ has a naive convergence range for all
$0< |x| <1$ which is defined to be the early universe with $a_0/e
< a < a_0$. Note that we are interested only in the past universe
where $a < a_0$. We will show in details how to extract the
leading terms in the $x$-expansion with $x=\ln (1+z)$ around the
point $x(z=0)=0$. One can write the expansion coefficient as
$\omega_n = {(1+\omega)^{(n)}(x=0) / n!}$ such that the power
series for the expansion of the EOS becomes
\begin{equation} \label{xx}
1+\omega(x(z))= \sum_n \omega_n x^n .
\end{equation}
Note that we use the same notation for $\omega_n$ in different
parameterizations for convenience. Hence the Eq. (\ref{eos}) can
be shown to be
\begin{eqnarray} \label{eos1x}
d \ln \rho (x)  = 3 (1+\omega)  d x= d\,\, \left[ \sum_n 3
\omega_n {x^{n+1} \over n+1} \right ] .
\end{eqnarray}
Note that we are now expanding the physical quantities with
respect to the function $(1+\omega)$ instead of $\omega$ for
convenience. Therefore, one can integrate above equation to obtain
\begin{equation} \label{eos2x}
\rho (x) = \rho_0 \exp [\,\, 3 \sum_n  {\omega_n x^{n+1} \over
n+1}\,\,] \equiv \rho_0 X(x)= \rho_0 \sum_n X_n x^n.
\end{equation}
Note that one needs to expand the function $\rho$ as a power
series of $x$ too in order to extract the approximated solution
with appropriate order. The expansion coefficient $X_n$ is defined
as $X_n = X^{(n)}(x=0)/n!$. One can show that $X'=XY$ with $Y=3
\sum_n  \omega_n x^n=3(1+\omega)$. Therefore, one has
\begin{equation}\label{xlx}
Y^{(l)}(x=0)=3(l!) \,\, \omega_l .
\end{equation}

One can also show, for example, that
\begin{eqnarray}
X''&=& X(Y^2+Y') \\
X'''&=& X(Y^3+3YY'+Y'') \\
X^{(4)} &=& X(Y^4+6Y^2Y'+3(Y')^2+4 YY''+Y''' ).
\end{eqnarray}
In addition, one can show that
\begin{equation}
X^{(l)}= \sum_n \sum_k 3 X_n \omega_k { (n+k)! \over (n+k-l+1)!}
x^{n+k-l+1}.
\end{equation}
Therefore, one has
\begin{equation}
X^{(l)}(0)= \sum_{n}^{l-1} 3 X_n \omega_{l-n-1} (l-1)!.
\end{equation}
Hence one obtains the recurrence relation for the expansion
coefficients of $X_n$:
\begin{equation}
X_l = {3 \over l} \sum_{n=o}^{l-1}  X_n \omega_{l-n-1} .
\end{equation}
As a result, one has, for example,
\begin{eqnarray}
X_0&=&1, \\
X_1&=& 3 \omega_0, \\
X_2&=&{1 \over 2} [9 \omega_0^2+3
\omega_1], \\
X_3&=& {1 \over 2} [9 \omega_0^3+9 \omega_0 \omega_1
+ 2 \omega_2 ], \\
X_4 &=& {1 \over 8} [27 \omega_0^4+54 \omega_0^2 \omega_1 + 9
\omega_1^2 + 24\omega_0 \omega_2+ 6\omega_3 ].
\end{eqnarray}
Therefore, one can expand the final expression for the energy
density $\rho$ accordingly. Indeed, one has
\begin{eqnarray}
\rho&=&\rho_0 \{  1 +  3 \omega_0 x + {1 \over 2} [9 \omega_0^2+3
\omega_1] x^2 +  {1 \over 2} [9 \omega_0^3+9 \omega_0 \omega_1 + 2
\omega_2 ] x^3  \nonumber \\ &+&
 {1 \over 8} [27 \omega_0^4+54 \omega_0^2 \omega_1 + 9
\omega_1^2 + 24\omega_0\omega_2+ 6\omega_3 ]x^4 \} +O(x^5)
\nonumber \\
&=& \rho_0 \left \{  \exp [3 \omega_0 x] + {3 \over 2} \omega_1
x^2 + {1 \over 2} [9 \omega_0 \omega_1
+ 2  \omega_2 ] x^3  
+ {1 \over 8} [54 \omega_0^2 \omega_1 + 9 \omega_1^2 + 24
\omega_0\omega_2+ 6\omega_3 ]x^4 \right \} +O(x^5).
\end{eqnarray}
In addition, one can show that the Hubble parameter $H=H_0X^{1/2}$
with $H_0=\sqrt{8\pi G \rho_0 /3}$. And the expansion for
$X^{1/2}$ can be obtained by replacing all $\omega_n$ with
$\omega_n/2$ in writing the expansion for $X$. The result is
\begin{eqnarray}
H&=& H_0 \left \{ \exp[ {3 \over 2}  \omega_0 x ] + {3 \over 4}
\omega_1 x^2 +  {1 \over 8} [9 \omega_0 \omega_1
+ 4  \omega_2 ] x^3  
+ {1 \over 32} [27 \omega_0^2 \omega_1 + 9 \omega_1^2 +
24\omega_0\omega_2+ 12\omega_3 ]x^4 \right \}+O(x^5).
\end{eqnarray}
Note also that one can also compute the conformal time according
to the expression:
\begin{equation}
H_0 \eta = \int_0^z dz'X^{-{1 \over 2}}=\int_0^x dx' X^{-{1 \over
2}} \exp [x'] =  \sum_n \int_0^x dx' X^{-{1 \over 2}} {x'^n \over
n !}.
\end{equation}
Therefore, one can easily compute the expansion of $\eta$ in a
straightforward manner. One can write $X^{-1/2}=\sum_k  x_k x^k=
\sum_k X_k(\omega_n \to -\omega_n/2) x^k$, and $H_0 \eta = \sum_k
\eta_k x^k$ for convenience. Therefore, one has
\begin{equation} \label{etaxx}
\eta_{l}= \sum_{n=0}^{l-1} {x_{l-n-1}  \over l \, n !}
\end{equation} for all $l \ge 1$.
Note that $\eta_0=0$.  As a result, one can easily reconstruct the
power series of $H_0 \eta$. Indeed, one has
\begin{eqnarray}
H_0 \eta &=&  x + {1 \over 4} (2- 3\omega_0 )x^2 +
 {1 \over 24} (9\omega_0^2 -6 \omega_1 -12\omega_0 + 4)x^3  \nonumber \\ &-&
 {1 \over 192} (27 \omega_0^3 -54  \omega_0^2  -54  \omega_0 \omega_1 +36\omega_0 +36 \omega_1
+ 24 \omega_2-8 )x^4  +O(x^5) \nonumber \\
 &=& {\rm e}^x -1- {3 \over 4} \omega_0 x^2 +
 {1 \over 8} (3\omega_0^2 -2\omega_1 -4 \omega_0 )x^3  
 -{1 \over 64} (9 \omega_0^3 -18  \omega_0^2  -18  \omega_0 \omega_1 +12\omega_0 +12 \omega_1
+ 8  \omega_2 )x^4  +O(x^5).
\end{eqnarray}
Note that a trivial complete sum of terms $x+x^2/2!+x^3/3!
\cdots={\rm e}^x-1$, derived from the $n=l-1$ terms in Eq.
(\ref{etaxx}), has been summed over for convenience in the final
expression.

\subsection{power series of $z $}

Note that the expansion parameter $z \equiv ( a_0/a)-1$. Hence the
range of $z$ of interest is the range $z<1$ in the past universe
which is equivalent to the range where $a_0/2 < a <a_0$.
Therefore, the Taylor series expansion around small $|z|$ is
equivalent to a series around $a=a_0$. This is again a series
expansion around the very recent universe near $a=a_0$. We will
show in details how to extract the leading terms in the
$z$-expansion around the point $z=0$. One can write the expansion
coefficient as $\omega_n = {(1+\omega)^{(n)}(z=0) / n!}$ such that
the power series for the expansion of the EOS becomes
\begin{equation} \label{zz}
1+\omega(z)= \sum_n \omega_n z^n .
\end{equation}
Note that we use the same notation for $\omega_n$ in different
parameterizations for convenience. Hence the Eq. (\ref{eos}) can
be shown to be
\begin{eqnarray} \label{eos1z}
d \ln \rho (z)  = 3 (1+\omega)  {d z \over 1+z}= d\,\, \left[
\sum_{n,k} (-1)^k 3 \omega_n {z^{n+k+1} \over n+k+1} \right ] .
\end{eqnarray}
Note that we are now expanding the physical quantities with
respect to the function $(1+\omega)$ instead of $\omega$ for
convenience. Therefore, one can integrate above equation to obtain
\begin{equation} \label{eos2z}
\rho (z) = \rho_0 \exp \left [\,\, 3 \sum_{n,k} (-1)^k {\omega_n
z^{n+k+1} \over n+k+1}\,\, \right ] \equiv \rho_0 X(z)= \rho_0
\sum_n X_n z^n.
\end{equation}
Note that one needs to expand the function $\rho$ as a power
series of $z$ too in order to extract the approximated solution
with appropriate order. The expansion coefficient $X_n$ is defined
as $X_n = X^{(n)}(z=0)/n!$. One can show that $X'=XY$ with $Y=3
\sum_{n,k}  (-1)^k \omega_n z^{n+k}$. Therefore, one has
\begin{equation}\label{Yl0}
Y^{(l)}(z)=3 \,\, \sum_{n} \sum_{k} (-1)^k { (n+k)! \over
(n+k-l)!} \omega_n z^{n+k-l}.
\end{equation}
Hence one has
\begin{equation}\label{Yl}
Y^{(l)}(z=0)=3 \;\; l! \,\, \sum_{n=0}^{l}  (-1)^{n+l}  \omega_n .
\end{equation}

One can also show, for example, that
\begin{eqnarray}
X''&=& X(Y^2+Y') \\
X'''&=& X(Y^3+3YY'+Y'') \\
X^{(4)} &=& X(Y^4+6Y^2Y'+3(Y')^2+4 YY''+Y''' ).
\end{eqnarray}
This series does not appear to have a more compact close form for
the multiple differentiation with respect to $z$. One can,
however, put the equations as a more compact format:
\begin{equation}
X^{(l+1)}=X [Y+{d \over dz}]^l Y.
\end{equation}
It appears, however, that one needs to do it manually even it is
straightforward. We will only list the leading terms as this is
already suitable expansion for our purpose at this moment.

Hence one has
\begin{eqnarray}
X_0&=&1, \\
X_1&=& 3 \omega_0, \\
X_2&=&{1 \over 2} [9 \omega_0^2-3\omega_0+3
\omega_1], \\
X_3&=& {1 \over 2} [9 \omega_0^3+9 \omega_0 (\omega_1-\omega_0)
+ 2 (\omega_0 - \omega_1 + \omega_2) ], \\
X_4 &=& {1 \over 8} [27 \omega_0^4+54 \omega_0^2
(\omega_1-\omega_0) + 9 (\omega_0 - \omega_1)^2 +
24\omega_0(\omega_0 - \omega_1 + \omega_2)+ 6(\omega_3 - \omega_2
+ \omega_1 -\omega_0) ].
\end{eqnarray}
Therefore, one can expand the final expression for the energy
density $\rho$ accordingly. Indeed, the result is
\begin{eqnarray}
\rho&=&\rho_0 \{  \exp [3 \omega_0 z ] - {3 \over 2} [\omega_0-
\omega_1] z^2 +  {1 \over 2} [9 \omega_0 (\omega_1-\omega_0)
+ 2 (\omega_0 - \omega_1 + \omega_2) ] z^3  \nonumber \\
&+& {1 \over 8} [54 \omega_0^2 (\omega_1-\omega_0) + 9 (\omega_0 -
\omega_1)^2 + 24\omega_0(\omega_0 - \omega_1 + \omega_2)+
6(\omega_3 - \omega_2 + \omega_1 -\omega_0) ]z^4  \} +O(z^5)
\end{eqnarray}
to the order of $z^4$. Note that one keep the order of precision
to $z^4$ in computing the energy density $\rho$ even we are
expanding the EOS only to the order of $z^3$. This is due to the
special structure in the energy momentum conservation law
(\ref{eos}). In addition, one can show that the Hubble parameter
$H=H_0X^{1/2}$ with $H_0=\sqrt{8\pi G \rho_0 /3}$. And the
expansion for $X^{1/2}$ can be obtained by replacing all
$\omega_n$ with $\omega_n/2$ in writing the expansion for $X$.
Therefore one has
\begin{eqnarray}
H&=&H_0  \{ \exp [ {3 \over 2} \omega_0 z ] + {3 \over 4}
(\omega_1- \omega_0) z^2 +  {1 \over 8} [9\omega_0
(\omega_1-\omega_0) +4 (\omega_0 - \omega_1 + \omega_2) ] z^3   \nonumber \\
&& + {1 \over 32}  [27 \omega_0^2 (\omega_1-\omega_0) +9 (\omega_0
- \omega_1)^2 + 24\omega_0(\omega_0 - \omega_1 + \omega_2)+
12(\omega_3 -\omega_2 + \omega_1 -\omega_0)  ] z^4  \} +O(z^5).
\end{eqnarray}

Note also that one can also compute the conformal time according
to the expression:
\begin{equation}
H_0 \eta = \int_0^z dz'X^{-{1 \over 2}}.
\end{equation}
One can write $X^{-1/2}=\sum_k  x_k z^k= \sum_k X_k(\omega_n \to
-\omega_n/2) z^k$ for convenience. Therefore, one has
\begin{equation}
H_0 \eta =\sum_{n=1}^\infty {x_{k-1} \over k} z^k
\end{equation}
As a result, one can easily reconstruct the power series of $H_0
\eta$. Therefore, one has
\begin{eqnarray}
H_0 \eta &=&  z - {3\omega_0 \over 4}  z^2 +
 {1 \over 8} [3\omega_0^2 + 2\omega_0 -2 \omega_1]z^3  
 -{1 \over 64} [9 \omega_0^3 + 18 \omega_0^2  -18 \omega_0 \omega_1 + 8 (\omega_0 -  \omega_1
+  \omega_2) ]z^4  +O(z^5) \nonumber \\
&=&  { 2 \over 3 \omega_0} ( 1- \exp[-{3\omega_0 \over 2}  z]\,\,)
+
 {1 \over 4} [\omega_0 - \omega_1]z^3  
 -{1 \over 16} [3 \omega_0^2  -3 \omega_0 \omega_1 + 4 (\omega_0 -  \omega_1
+  \omega_2) ]z^4  +O(z^5).
\end{eqnarray}

\section{Comparison and Analysis}
In summary, one has obtained the series sum of the corresponding
energy density, Hubble parameter and conformal time for different
parameterizations of the function $\omega$. For $y=z/(1+z)$ one
has:
\begin{eqnarray}
\rho(y) &=& \rho_0 \{ \exp [3 \omega_0 y] + {3 \over 2} [\omega_0+
\omega_1] y^2 +  {1 \over 2} [9 \omega_0 (\omega_0+\omega_1)
+ 2 (\omega_0 + \omega_1 + \omega_2) ] y^3  \nonumber \\
&+& {1 \over 8} [54 \omega_0^2 (\omega_0+\omega_1) + 9 (\omega_0 +
\omega_1)^2 + 24\omega_0(\omega_0 + \omega_1 + \omega_2)+
6(\omega_0 + \omega_1 + \omega_2 +\omega_3) ]y^4  \} +O(y^5)
\\
H(y)&=&H_0  \{  \exp[{3 \over 2} \omega_0 y] + {3 \over 4}
(\omega_0+ \omega_1) y^2 +  {1 \over 8} [9 \omega_0
(\omega_0+\omega_1)
+ 4 (\omega_0 + \omega_1 + \omega_2) ] y^3 + {1 \over 32}  [  \nonumber \\
&& 27 \omega_0^2 (\omega_0+\omega_1) +9 (\omega_0 + \omega_1)^2 +
24\omega_0(\omega_0 + \omega_1 + \omega_2)+ 12(\omega_0 + \omega_1
+ \omega_2 +\omega_3)  ] y^4  \} +O(y^5).
\\
H_0 \eta (y)
 &=& {y \over 1-y} - {3 \over 4} \omega_0 y^2 + [
 {3 \over 8} \omega_0^2 -{1 \over 4} \omega_1 -{5 \over 4} \omega_0 ]y^3
 -[{9 \over 64} \omega_0^3- {27 \over 32} \omega_0^2  -{9 \over
32} \omega_0 \omega_1 +{13 \over 8} \omega_0 +{1 \over 2} \omega_1
+ {1 \over 8} \omega_2 ]y^4 +O(y^5).
\end{eqnarray}
For $x=\ln (1+z)$, one has:
\begin{eqnarray}
\rho(x) &=& \rho_0 \left \{  \exp [3 \omega_0 x] + {3 \over 2}
\omega_1 x^2 + {1 \over 2} [9 \omega_0 \omega_1
+ 2  \omega_2 ] x^3  
+ {1 \over 8} [54 \omega_0^2 \omega_1 + 9 \omega_1^2 + 24
\omega_0\omega_2+ 6\omega_3 ]x^4 \right \} +O(x^5).
\\
H (x)&=& H_0 \left \{ \exp[ {3 \over 2}  \omega_0 x ] + {3 \over
4} \omega_1 x^2 +  {1 \over 8} [9 \omega_0 \omega_1
+ 4  \omega_2 ] x^3  
+ {1 \over 32} [27 \omega_0^2 \omega_1 + 9 \omega_1^2 +
24\omega_0\omega_2+ 12\omega_3 ]x^4 \right \}+O(x^5).
\\
H_0 \eta(x)
 &=& {\rm e}^x -1- {3 \over 4} \omega_0 x^2 +
 {1 \over 8} (3\omega_0^2 -2\omega_1 -4 \omega_0 )x^3  
 -{1 \over 64} (9 \omega_0^3 -18  \omega_0^2  -18  \omega_0 \omega_1 +12\omega_0 +12 \omega_1
+ 8  \omega_2 )x^4  +O(x^5).
\end{eqnarray}
For the $z$ expansion, one has:
\begin{eqnarray}
\rho(z)&=&\rho_0 \{  \exp [3 \omega_0 z ] - {3 \over 2} [\omega_0-
\omega_1] z^2 +  {1 \over 2} [9 \omega_0 (\omega_1-\omega_0)
+ 2 (\omega_0 - \omega_1 + \omega_2) ] z^3  \nonumber \\
&+& {1 \over 8} [54 \omega_0^2 (\omega_1-\omega_0) + 9 (\omega_0 -
\omega_1)^2 + 24\omega_0(\omega_0 - \omega_1 + \omega_2)+
6(\omega_3 - \omega_2 + \omega_1 -\omega_0) ]z^4  \} +O(z^5)
\\
H(z)&=&H_0  \{ \exp [ {3 \over 2} \omega_0 z ] + {3 \over 4}
(\omega_1- \omega_0) z^2 +  {1 \over 8} [9\omega_0
(\omega_1-\omega_0) +4 (\omega_0 - \omega_1 + \omega_2) ] z^3   \nonumber \\
&& + {1 \over 32}  [27 \omega_0^2 (\omega_1-\omega_0) +9 (\omega_0
- \omega_1)^2 + 24\omega_0(\omega_0 - \omega_1 + \omega_2)+
12(\omega_3 -\omega_2 + \omega_1 -\omega_0)  ] z^4  \} +O(z^5). \\
H_0 \eta(z) &=&  { 2 \over 3 \omega_0} ( 1- \exp[-{3\omega_0 \over
2} z]\,\,) +
 {1 \over 4} [\omega_0 - \omega_1]z^3  
 -{1 \over 16} [3 \omega_0^2  -3 \omega_0 \omega_1 + 4 (\omega_0 -  \omega_1
+  \omega_2) ]z^4  +O(z^5).
\end{eqnarray}

Note that the expansion coefficients $\omega_n$ are defined
differently for different expansions even we are using the same
notation for convenience. Also note that at small red-shift $z \ll
1$, $y$ and $x$ are both very close to $z$. This is the reason why
we are ending up with similar leading term in each expansion. Once
the red-shift $z$ is extended, difference in the leading terms
will be significant. One is therefore able to distinguish the
contribution of the next-leading terms.

As mentioned earlier, one should be able to determine the
expanding coefficients $\omega_n$ with the result from the
measurements of future experiments. Once the expansion
coefficients are determined from fitting with the Hubble parameter
or energy density, one will be able to reconstruct the function
$\omega$ to higher order and higher precision. The result should
be helpful to determine the origin and the nature of the matter
sources and to be compared with some possible combinations of
various fundamental theories.

The linear parametrization of the $\omega$ has been applied to the
study of the evolution history of our universe. The linear models
adopted has been shown to be useful in making predictions for
future observations \cite{[13]}. One is naturally lead to answer
the question whether one should take the linear model as a
complete theory and integrate it without worrying about the higher
order corrections. To be more specific, linear model like
$\omega(z)= \omega_0'+\omega_1 z$ has been taken as a complete
theory. Related physical quantities are computed with all power of
$\omega_1$, in the power series shown above, summed over for the
final result. On the other hand, if the linear model is taken as
leading order approximation, one should ignore higher order
contributions from $z^n ( \forall  n \ge 2)$.

If higher order corrections is small and negligible, it does not
matter much whether one ignores or includes all higher order (in
$\omega_1$) corrections. For example, in a convergent series,
higher order contributions are smaller and smaller order by order.
Therefore, it is fine to sum over all terms related to the
expansion parameter $\omega_1$. Otherwise, one should pay
attention to the convergent properties of these expansions for
possible deviations.

Looking at our results, it is easy to find that a re-summation of
the expansion coefficients $\omega_n$ in these formulae is
difficult to obtain except the re-summation of $\omega_0$ which
has been already shown in above equations. Fortunately, one can do
it by a different way which will be shown in a moment. As a
result, one would be able to see whether an expansion in power of
$\omega_n$ is reliable or not.

In next section, we will show how to sum up all terms order by
order as series of expansion coefficients $\omega_n$. Taking the
linear model $\omega=\omega_0'+\omega_1 y$ as a complete theory
and include all effect of $\omega_0'$ and $\omega_1$ include more
effect from the higher order $y^n$ contribution as compared to the
leading order approximation approach. The inclusion of these
effect may not be reliable unless they are small as compared to
the leading terms.

In addition, taking linear model as a complete theory, the effects
from  all higher order in $\omega_{n>1}$ are ignored. The
truncation may need to appreciable error in evaluation of related
physical observables unless these higher order terms are also
small. We will also analyze the error due to the contributions
from the higher order $\omega_{n>1}$ term.

Note again that linear model may serve as a good approximation
theory in $y$ expansion because the naive convergence range $y<1$
covers the entire history of our universe from $z=0 \to \infty$.
If the linear model fits the original theory very well at small
$y$ and higher order contributions are not appreciable, linear
model can be extended to remain valid even at larger $y$. The
higher order effect related to $\omega_2$ will be evaluated also
for the hope that future experiments may provide better resolution
to distinguish possible minor effect.

In addition, one more advantage of the linear model written as
$\omega=\omega_0'+\omega_1 y$ over the linear model in expansion
of $z$ can be easily seen from the relation:
\begin{equation}
\omega(y)=\omega_0'+\omega_1 y=\omega_0'+\omega_1 {z \over
1+z}=\omega_0'+\omega_1 z(1-z+z^2- \cdots)=\omega(z)-\omega_1
z^2(1-z+z^2+ \cdots).
\end{equation}
Indeed, one can see that linear model $\omega(y)$ contains a few
more higher order $z^{n>1}$ effect than the linear model
$\omega(z)\equiv \omega_0'+\omega_1 z$. The effect of the higher
order terms depends, however, on the actual deviation from the
underlying theory like SUGRA or inverse power law models.
Evidences show that linear model in $y$ works better than the
linear model in $z$ at large redshift for both reasons.

\section{error estimation of the linear model }
Note that the expansion parameter $y \equiv z/(1+z)= 1 -a/a_0$.
Therefore, the Taylor series expansion around small $y$ is
equivalent to a series expansion around $a=a_0$. This is again a
series expansion around the very recent universe near $a=a_0$. The
expansion series in $y$ has a naive convergence range for all $y
<1$ which is defined to be the complete early universe with $a <
a_0$. We will show in details how to extract the leading terms in
the $y$-expansion with $y={z / (1+z})$ around the point
$y(z=0)=0$.

Indeed, one can write the expansion coefficient as $\omega_n =
{(1+\omega)^{(n)}(y=0) / n!}$ such that the power series for the
expansion of the EOS becomes $1+\omega(y(z))= \sum_n \omega_n y^n
$. One can write $w=1-y$ and write it as
\begin{equation} \label{yw}
1+\omega(y)= \sum_n \omega_n y^n =\omega_0+ \sum_{n =1}^\infty
\omega_n (1-w)^n= \omega(0)+ \sum_{n =1}^\infty\sum_{k =1}^n C^n_k
\omega_n(-w)^k .
\end{equation}
Here $\omega(0)=\sum_{n =0}^\infty \omega_n=[1+\omega(y)]_{y=0}$.
Hence the Eq. (\ref{eos}) can be shown to be
\begin{eqnarray} \label{eos2}
d \ln \rho (y)  = 3 (1+\omega)  { d y \over 1-y}= - 3 (1+\omega) {
d w \over w}= -3 d \left[ \omega(0) \ln w + \sum_{n
=1}^\infty\sum_{k =1}^n C^n_k { \omega_n (-w)^k \over k}  \right
].
\end{eqnarray}
Therefore, one can show that
\begin{equation}\label{rho20}
\rho(y) = \rho(0) \left [ (1-y)^{-3 \omega(o)} \exp [ -3 \sum_{n
=1}^\infty\sum_{k =1}^n \sum_{l =1}^k \; C^n_k C^k_l { \omega_n
y^l \over k}   ] \right ].
\end{equation}
In fact, one can show that when $y \to 1$, the energy density
shown in Eq. (\ref{rho20}) is dominated and proportional to
\begin{equation}
\rho (y \to 1) \propto (1-y)^{-3 \omega(o)}
\end{equation}
implying $\rho \propto (1-y)^{-3(0.18+\omega_2+\omega_3+ \cdots)}$
for the model approximated by $\omega=-0.82+0.58y$. This term will
diverge at $y=1$ if all higher order coefficients are small as
compared to the leading coefficients such that the sum $\omega(0)$
remain positive. Therefore, one can expand the density as a power
series of $\omega_n$ from above equation. For comparison, one can
show that
\begin{equation} \label{rho2}
\rho_2(y) = \rho_1(y) \left [ (1-y) \exp [ y+{y^2 \over 2} ]
\right ]^{-3 \omega_2}
\end{equation}
keeping terms to the order of $\omega_2$. Here $\rho_1(y)$ denotes
$\rho(y)$ with $\omega_n=0$ for all $n \ge 2$. Similarly,
$\rho_2(y)$ denotes $\rho(y)$ with $\omega_n=0$ for all $n \ge 3$.
Hence the last factor of above equation provides the $\omega_2$
corrections of the power series.

Note that linear model $\omega=-0.82+0.58y$ is considered as a
complete model with the corresponding energy density $\rho=\rho_1$
shown in Eq. (\ref{rho2}). As a complete model, all powers of
$y^n$ are included in the evaluation of energy density $\rho_1$.
Indeed, one can show that
\begin{eqnarray}
&& \rho_1=\rho_0(1-y)^{-3(\omega_0+\omega_1)} \exp [ -3 \omega_1
y], \\
&& \rho^1= \rho_0 \{  1 +  3 \omega_0 y + {1 \over 2} [9
\omega_0^2+3\omega_0+3 \omega_1] y^2 \}
\end{eqnarray}
with $\rho^1$ given above denoting the truncated energy density to
the order of $y^2$ for the approximated counterpart. Results shown
in Fig. 1 indicates that the difference between $\rho_1$ and
$\rho^1$ is small when $y$ is small. But it becomes appreciable at
larger $y$. Note again that $\omega_0 \equiv 1+\omega(y=0)=0.18$
in our notation for the linear model $\omega=-0.82+0.58y$.
\begin{figure*}
\includegraphics{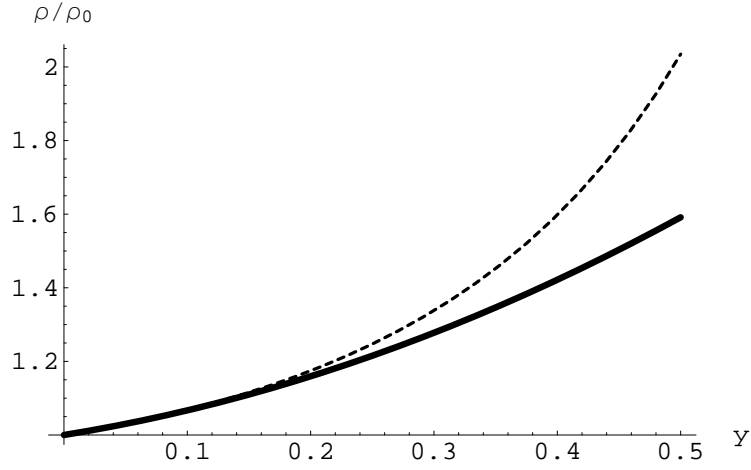}
\caption{\label{fig:rho} Thin dotted line plots $\rho_1(y)/\rho_0$
and thick solid lines represents $\rho^1(y)/\rho_0$ for the linear
model $\omega=-0.82+0.58y$. }
\end{figure*}

Similarly, one can show that $H^2 = 8 \pi G \rho/3$ can be
integrated to give:
\begin{equation}
H(y) = H(0) \left [ (1-y)^{-3 \omega(o)/2} \exp [ -{3 \over 2}
\sum_{n =1}^\infty\sum_{k =1}^n \sum_{l =1}^k \; C^n_k C^k_l {
\omega_n y^l \over k}   ] \right ].
\end{equation}
Therefore, one can also expand the density as a power series of
$\omega_n$ from above equation. Indeed, one can show that
\begin{equation}
H_2(y) = H_1(y) \left [ (1-y) \exp [ y+{y^2 \over 2} ] \right
]^{-3 \omega_2/2}.
\end{equation}
Here $H_1(y)$ denotes $H(y)$ with $\omega_n=0$ for all $n \ge 2$.
Similarly, $H_2(y)$ denotes $H(y)$ with $\omega_n=0$ for all $n
\ge 3$. Hence the last factor of above equation provides the
$\omega_2$ corrections of the power series.

\begin{figure*}
\includegraphics{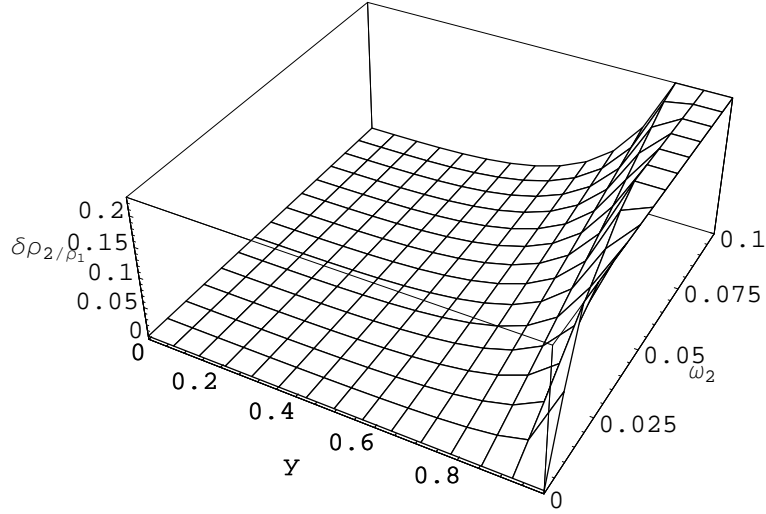}
\caption{\label{fig:rho} The error $\rho_2/\rho_1-1$ due to the
presence of the $\omega_2$ correction, as compared to the linear
model, is plotted in 3D format as a function of $y$ and
$\omega_2$.}
\end{figure*}

In Fig. 2, the error $\rho_2/\rho_1-1$ due to the presence of the
$\omega_2$ correction, as compared to the linear model, is plotted
in 3D format as a function of $y$ and $\omega_2$. It is easy to
see that the error becomes appreciable when $y >7$ and $\omega_2 >
0.4$.

\begin{figure*}
\includegraphics{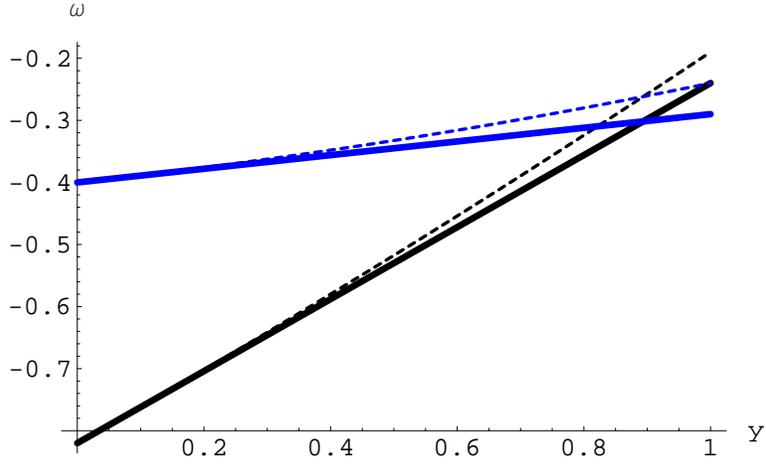}
\caption{\label{fig:omega} The equation of state $\omega$ is
plotted as a function of $y$. Lower thick solid line denotes the
contribution of linear model $\omega=-0.82+0.58y$. Lower thin
dotted line represents the addition of the $y^2$ correction with
$\omega_2 y^2=0.1y^2$. Similarly, Upper thick solid line denotes
the contribution of linear model $\omega=-0.4+0.11y$. Upper thin
dotted line represents the corresponding $y^2$ correction from
$\omega_2 y^2=0.1y^2$. }
\end{figure*}
In Fig. 3, the equation of state $\omega$ is plotted as a function
of $y$. Lower thick solid line denotes the contribution of linear
model $\omega=-0.82+0.58y$\cite{[15]}. This linear model fits
SUGRA equation of state very well at small $y$. Lower thin dotted
line represents the addition of the $y^2$ correction with
$\omega_2 y^2=0.1y^2$. In addition, the upper thick solid line
denotes the contribution of linear model
$\omega=-0.4+0.11y$\cite{[15]}. This model fits the Inverse power
law model very well at small $y$. Upper thin dotted line
represents the corresponding $y^2$ correction from $\omega_2
y^2=0.1y^2$. Note that it has been shown from Figure 1 of
Reference \cite{[15]}, linear models shown above fits the SUGRA
and inverse power law model very well at small $y$ as compared to
the linear model in $z$.
$\omega$ as a

It is easily seen that a small $y^2$ correction with
$\omega_2=0.1$ does not modify the function $\omega$ very much. In
addition, the correction to the density function for
$\omega_2=0.1$ is not appreciable according to Fig. 2. Therefore,
the linear model can be truncated at the order of $\omega_1$ for
small $y$ without affecting the final result in an appreciable
way.

From above presentations, one also finds that there are two
comparisons required for the reliability for the linear model.
First of all, one should check if the contributions from higher
order coefficients $\omega_{n >1}$ will affect the evaluation
results. Secondly, one should also check if the inclusion of
higher order terms $y^{n>2}$ in the physical functions, e.g.
$\rho_1$, will affect the final result in an appreciable way.

Due to the fact that the naive convergence range of $y=z/(1+z)$
expansion covers the entire history of our early universe, this
expansion appears to be the best way to approximate our physical
models. As a result, the approximated linear model may represent
the complete theory effectively for a larger domain $y$ provided
that the higher order corrections derived from $\omega_2$ is
small. Our results indicates that this appears to be the case for
the SUGRA models and inverse power law models\cite{[15]}.
Therefore, the corresponding linear model appears to be more
reliable than the other linear models expanded in $z$ or $\ln
(1+z)$.

Detailed analysis indicates that the linear model is a very
reliable approximation within small $y$. Beyond the region of
small $y$, one should be very careful with the higher order
contributions when comparing with the future experiments. In
practice, the sources of equation of state $\omega$ may be very
complicate. In spite of the fact that it may be derived from
complicate combinations of many different sources, linear model
provides an easy way to determine the leading coefficients that
should be very reliable at small $y$ corresponding to our very
recent universe. If one is able to determine the next leading term
coefficients $\omega_2$ from better precision measurements in the
near future, it would provide better way to distinguish the nature
of the source of equation of state.

\section{conclusion}

Linear models, e.g. $\omega=\omega_0'+\omega_1 y$, should be
thought of as leading approximation of a complete Taylor series.
If this is so, result derived from higher order in $y$, e.g.
$O(y^2)$ terms, should only provide marginal contribution and
hence can be ignored. If the higher order contribution is
appreciable, one should be very careful in dealing with the
truncated series. As a practical analysis, one should compute all
related physical functions also as a power series order by order
to obtain reliable observable according to the precision
requirement. Otherwise, unphysical contributions may build up and
leave the final result invalid. Fortunately, this complication may
be unnecessary if the higher order contributions are small as
compared to the leading order contribution.

Indeed, linear models of equation of state are considered as
effective models for some underlying theoretical models. Many
theories including Sugra and Inverse power law model can
approximated by linear model described by
$\omega=\omega_0'+\omega_1 y$. Therefore these linear models are
potentially good candidate to simulate the large redshift
properties of the underlying theories because the higher order
corrections could be negligible at larger $z$, corresponding to $y
\equiv z/(1+z) \to 1$.

Since $z \to \infty$ is equivalent to $y \to 1$. Hence y-expansion
with $y \to 1$ is still in the range of naive convergence of the
corresponding Taylor series. Therefore, as long as the higher
order terms are negligible as compared to the liner approximation,
one is free to consider the linear model as a well-behaved
representation of the underlying theory. Otherwise, one should pay
attention to possible deviations derived from the higher order
corrections.

We have calculated two possible errors when one treats the linear
model as a complete theory instead of the leading approximation
for an underlying theory. One possible error comes from the
inclusion of all higher order terms (in $y^n$ ) related to
$\omega_0$ and $\omega_1$ in the linear model calculation. The
other error comes from the truncated higher order terms in
$\omega_n$ for the underlying theory. They are both computed and
compared in section V.

We have also tried to analyze possible deviations for the Sugra
and inverse power law models in this section. Results show that
linear model $\omega=\omega_0'+\omega_1 y$ for these two models
appears to be a good approximation even at large redshift region
as long as the higher order corrections are small.

Since we are only able to determine only a few leading
coefficients in the future experiments, linear model appears to
serve this purpose very well. Hopefully, one should, however,  be
able to determine for example the deviation due to the $\omega_2$
contributions and further the understanding of the underlying
theory if better resolution can be made possible in the future
experiments .

In addition, we have also tried to provide a more complete list of
the Taylor series in this paper. Hope that these presentations can
provide better information for the quest of the mapping of our
early universe.

Therefore, we try to extend the parametrization further to include
the effect of higher orders of the series expansion results.
Related physical quantities are treated carefully order by order
in order to extract more reliable information from these
expansions. One also tries to determine the expansion coefficients
from the fitting of the measured Hubble parameters or energy
density. As a result, one may reconstruct the series expansion of
the equation of state and probe the nature and origin of the
matter sources.

As mentioned above, we also present an error analysis to find the
range of convergence and possible error control for a meaningful
truncation. The linear model given by $\omega=-0.82+0.58 y$ is
known to fit the SUGRA result to a good precision at large $z$
even the linear model is already off $27 \%$ when $z \sim 1.7$.
One shows that the next leading expansion coefficient with
$\omega_2y^2 <0.1y^2$ only provides at most $20 \%$ error to the
density function even when the redshift is close to 1100. The
error is even down to $10 \%$ for the Hubble function $H$. As a
result, one show that a reliable truncation is possible when the
higher order expansion coefficient is small for all range
$y=z/(z+1)=0 \to 0.9991$ up to the last scattering surface when $z
\sim 1100$. Therefore, it does not matter much whether one sums up
all power of $y$ coupled to the leading coefficients $\omega_0$
and $\omega_1$ as long as the higher order corrections due to
$\omega_2$ and $\omega_{n>2}$ are negligible.

In spite of the fact that equation of state $\omega$ may be
derived from complicate combinations of many different sources,
linear model provides an easy way to determine the leading
coefficients that should be very reliable at small $y$
corresponding to our very recent universe. If one is able to
determine the next leading term coefficients $\omega_2$ from
better precision measurements in the near future, it would provide
better way to distinguish the nature of the source of equation of
state.

Indeed, the equation of state can be described by
$\omega=p/\rho=(\dot\phi^2/2 -V)/(\dot\phi^2/2 +V)$ for many
different effective theory with effective potential $V$ and scalar
field $\phi$. For example, $V_{INV} \sim \phi^{- \alpha}$ and
$V_{SUGRA} \sim \phi^{- \alpha} \exp [\phi^2]$ for inverse power
law models and SUGRA models respectively\cite{[15]}. The nature of
our physical universe may also consist of many other combinations
of scalar fields contributions. It is therefore important for us
to figure out which model plays the most important role in the
expansion history of our universe. The linear model or leading
approximation appears to be of help in determining the essential
part of the nature of the equation of state and its corresponding
effective theory. Therefore a more detailed analysis on the
convergence properties of various approximated models is very
important as a research topic.

In summary, one has evaluated the series expansion for the energy
density, Hubble constant and the conformal time for the $y$, $x$
and $z$ expansion of the corresponding equation of state function
$\omega$ up to order four in previous section. The proposed
Supernova/Acceleration Probe (SNAP) will carry out observations
aiming to determine the equations of state of the energy density,
providing insights into the cosmological model, the nature of the
accelerating dark energy, and potential clues to fundamental high
energy physics theories and gravitation. As a result, we show all
suitable ways to parameterizing the equation of state for
application to study its effect on the expansion history of the
recent universe.

A detailed discussion on the choices of the expansion parameters
for the Taylor series of the equation of states $\omega$ is
presented in this paper accordingly. For example, the Taylor
series of the EOS is expanded as power series of the variables
$y=z/(1+z)$, $x=\ln (1+z)$ and $z$ respectively. Due to the fact
that the naive convergence range of $y=z/(1+z)$ expansion covers
the entire history of our early universe, this expansion appears
to be the best way to approximate our physical models. As a
result, the approximated linear model may represent the complete
theory effectively for a larger domain $y$ provided that the
higher order corrections derived from $\omega_2$ is small. This
appears to be the case for the SUGRA models and inverse power law
models. Therefore, the corresponding linear model appears to be
more reliable than the other linear models expanded in $z$ or $\ln
(1+z)$.

We also show how to obtain the power series for the energy density
function, the Hubble parameter and related physical quantities of
interest. The method presented here will have significant
application in the precision distance-redshift observations aimed
to map out the recent expansion history of the universe, including
the present acceleration and the transition to matter dominated
deceleration.

Since we can power-expand all smooth EOS into a convergent power
series for a reasonable range of the expansion parameters, it is
more practical for the future probe to determine the expansion
coefficients $\omega_n$ or equivalently the local derivatives of
the EOS. One may need to use different expansion series depending
on the convergent speed of the power series. For example, it
appears that the leading order term in the $y$ expansion is better
as a nice result for the SUGRA prediction. This is because the
leading term is close enough to the theoretical prediction at
small $y$ region\cite{[13]}. Nonetheless, one expects that fitting
for a few more leading terms in the Taylor series will be able to
provide us better information about the nature of the function
$\omega$. In addition, the result shown here is independent of the
choice of the time $t_0$. The local measurement of the expansion
coefficients can be extended to the comparison of the expansion
coefficients at any time.

\section*{Acknowledgments}

This work is supported in part by the National Science Council of
Taiwan.

\end{document}